# Electric-field-induced metal maintained by current of the Mott insulator $Ca_2RuO_4$

Fumihiko Nakamura[1]\*, Mariko Sakaki[1], Yuya Yamanaka[1], Sho Tamaru[1], Takashi Suzuki[1] and Yoshiteru Maeno[2],

ADSM, Hiroshima University, Higashi-Hiroshima 739-8530, Japan[1],
Department of Physics, Kyoto University, Kyoto 606-8502, Japan[2].

Recently, "application of electric field (*E*-field)" has received considerable attention as a new method to induce novel quantum phenomena since application of *E*-field can tune the electronic states directly with obvious scientific and industrial advantages over other turning methods. However, *E*-field-induced Mott transitions are rare and typically require high *E*-field and low temperature. Here we report that the multiband Mott insulator $Ca_2RuO_4$ shows unique insulator-metal switching induced by applying a dry-battery level voltage at room temperature. The threshold field $E_{th}$ ~40 V/cm is much weaker than the Mott gap energy. Moreover, the switching is accompanied by a bulk structural transition. Perhaps the most peculiar of the present findings is that the induced metal can be maintained to low temperature by a weak current.

Correspondence to fumihiko@hiroshima-u.ac.jp

In the last few decades, there has been growing interest in developing energy-saving devices based on novel quantum phenomena. In particular, the Mott transition is one of the most attractive many-body effects [1]. Pressure ($P$) is a suitable tuning method to explore such novel phenomena; however, high-$P$ conditions are generally achieved in a complicated apparatus requiring sophisticated skills. In contrast, there has been growing recognition of applying $E$-field as a complementary method to $P$, since an insulator-metal switching, namely controlling $E$-field, has many advantages for practical use, especially for electronic devices. Moreover, application of $E$-fields can govern the electronic states directly whereas $P$ tunes the electronic states indirectly via the change of lattice parameters.

Let us mention two well-known examples of $E$-field-induced phenomena. One is "electrostatic carrier doping" (ESD) [2, 3], which controls carrier density in a surface region of an insulator by an extremely large electric field. Indeed, $E$-field-induced superconductivity has recently been reported in the surface layer of a band insulator $SrTiO_3$ by the $E$-field application of 20 kV/cm [4, 5]. The other is a Mott transition induced by applying $E$-fields, namely, "switching". Some transition-metal oxides in strongly correlated electron systems have gained attention as a candidate for the switching to practical use as an electronic device such as resistance RAM [6, 7]. Most of these previously reported switching phenomena have been achieved only at low temperatures and/or high voltage (typically 1 - 100 kV/cm) [8-12]. In order to develop energy saving devices, it is essential to find a switching system driven at room temperature (RT) and by weak $E$-fields. An example is the metal-insulator-transition (MIT) induced above room temperature in films of a Mott insulator $VO_2$ [8]. It would be desirable to find a system in which bulk metallic state is induced by electric switching with low $E$-field.

As another challenge, it is desirable to maintain the $E$-field-induced metallic state in a "volatile" switching system on cooling and identify interesting ground states, since there have been few such reports in a system stabilising a steady but nonequilibrium state.

We have devoted considerable efforts on the Mott insulator $Ca_2RuO_4$ (CRO) to induce switching and explore its ground state because CRO has the following versatile properties. Firstly, pressurised CRO displays a variety of quantum states, ranging from an antiferromagnetic (AFM) Mott insulator to superconductivity via a ferromagnetic (FM) quasi-two-dimensional metal [13-15]. Secondly, the magnetic and electronic properties of CRO are known to be quite sensitive to coupling of spin, charge and the orbital degrees of freedom [16, 17]. Lastly, the metalisation of the Mott insulator CRO can be achieved by heating above $T_{MIT}$= 357 K [18].

The gap energy in CRO is 0.2 eV determined from the activation energy based on the temperature ($T$) dependence of resistivity [19]. On the basis of a simple Zener breakdown model [20] we estimate $E_{th}$ for CRO to be ~4 MV/cm (we use here the in-plane lattice spacing of $a$ = 5.45 Å. Since $\varepsilon_F$ has been unknown for CRO, the Fermi energy of $\varepsilon_F$ ~0.2 eV for the $\gamma$ band of $Sr_2RuO_4$ from ref. 21).

**Results**

To investigate the switching phenomena, voltage-current ($V$-$I$) curves have been measured by using a two-probe method for CRO single crystals. Let us first present the results of *V*-biased experiments. Figure 1 (a) shows changes in *I* at 295 K as a function of *V*. With increasing *V* along the *c* axis, *I* first rises linearly at a rate indicating nonmetallic conduction of ~60 Ωcm, but then jumps discontinuously from 18 to 700 mA at 0.8 V, indicating switching, and is followed by an increase at a rate indicating metallic conduction of ~0.4 Ωcm. Surprisingly, the threshold value $E_{th}$ ~40 V/cm for $E_{\|c}$ is far smaller than our expectation of ~4 MV/cm. We typically obtained $E_{th}$~50 V/cm for $E_{\perp c}$. Thus, the value of the $E_{th}$ is almost independent of the *E*-field direction.

With reducing *V*, *I* decreases with the metallic slope. However, *I* vanishes abruptly at ~10 V/cm because the sample breaks into pieces (single crystalline CRO disintegrates not in the process of the insulator-to-metal transition but in the metal-to-insulator transition. Thus, the *E*-field induced disintegration occurs only in the decreasing *V* process at $E < E_{th}$). Until this disintegration occurs, the *V-I* curves show a large hysteresis indicating a first order Mott transition [22, 23] during *V* sweeps.

We show that simple Joule heating is negligible from the following three pieces of experimental evidence: Firstly, there was no appreciable change in *T* of the sample during *V* sweeps; secondly, the switching is also induced by applying only one tiny electric pulse such as $V_{th}$ ~6 V and *I* ~20 mA for the duration of 100 ms. In this switching, the total power of ~12 mJ is fed into a sample in contact with a heat bath. Even if the sample absorbs all the heat, the possible *T* rise of the sample (2.5 mg) is less than ~7 K. Thus, the actual *T* should remain much less than $T_{MIT}$ =357 K. Lastly, the *IV* curves obtained for different duration time are shown in Fig. 1(b) in order to characterize the amount of heat needed to induce switching. Total heating $Q_{th}$ at a threshold point is plotted as a function of the duration time in Fig. 1(c). $Q_{th}$ rises almost linearly with the duration time, in sharp contrast to constant $Q_{th}$ expected for a heating-dominated case. Thus, our switching phenomena cannot be interpreted in terms of a Joule heating.

From the V-I curves, we obtained the threshold values $E_{th}$ at several temperatures below 320 K. Figure 1(d) shows the $E_{th}$ divided by $E_0$, the extrapolated value of the $E_{th}$ to absolute zero, plotted as a function T. The $E_{th}$ rises on cooling. The linear relation in a logarithmic scale is characteristic of T variation of the $E_{th}$. The linear line is a fit with a formula of $E_{th}(T)/E_0 = \exp(-T/T_0)$, using the characteristic values $E_0$ = 80 kV/cm and $T_0$ = 39.2 K.

We, next, consider whether the switching in CRO occurs in local or in bulk. As discussed in the development of switching into the resistance RAM, many of switching phenomena in insulating oxides have successfully been interpreted in terms of local switching. Two kinds of local-switching models have recently been proposed: one is due to formation of a filamentary path as seen in highly insulating oxides such as NiO [24], and the other is due to interface-resistance switching as seen in relatively conductive perovskite-oxide insulators [25]. In the case that switching occurs locally, the size effect on resistance and threshold voltage should be different from that in a bulk switching case. To inspect this point, we examine the V-I curves obtained by using a four-probe method for the samples with different sizes and shapes.

To clarify whether the switching is due to a filamentary path or not, we used single-crystals of CRO formed into a shape with a step with different cross-sectional area. Current is plotted as a function of voltage and E-field in Fig.

1 (e) and (f), respectively. In both plots, a difference is visible between large and small cross-sectional area. In contrast, the plots of current density, $J$, against $E$-fields in Fig. 1(g) are universal behavior of the switching in a wide range of $E$-field. From the initial linear slope in the $J$-$E$ curve the resistivity of ~4.7 $\Omega$cm is obtained and agrees well with the resistivity obtained by an ac measurement. Moreover, the value of 4.7 $\Omega$cm indicates that the system is weakly conductive in bulk at RT. Thus, we conclude that the switching in CRO is not due to a filamentary path.

Next, we examine another possibility due to the interface resistance switching. In this case, the switching probability should increase with the area of electrodes because the switching occurs on the interface between a sample and an electrode. We have, however, observed that the resistance and the threshold voltage are almost independent of the area of electrodes. Moreover, the threshold voltage rises lineally with the distance of the electrodes. These clearly indicate that the switching in CRO is not due to interface resistance switching (many of switching phenomena in oxides can successfully be interpreted in terms of diffusion of oxygen defects; however, this model is not suitable for the switching in CRO because CRO does not prefer to produce the oxygen defects).

Now we examine the relation between structure and electronic properties in single-layered ruthenates. It has been known that the electronic phase stability of CRO are governed not simply by the effective correlation energy $U/W$, but also by the orbital degeneracy of the Ru$^{4+}$ $t_{2g}$ levels, both of which may abruptly change due to the RuO$_6$-octahedral distortions of flattening, tilt and rotation [22, 23]. In particular, the Mott transition is mainly due to the Jahn-Teller effect which produces a change in the orbital occupation associated with the flattening distortion [16].

In pure CRO, the high-$T$ ($T$ > 357 K) or high-$P$ ($P$ > 0.5 GPa) metallic phase shows the structure called "L-Pbca", with a weaker flattening and as well as a weaker tilt and smaller volume than the low-$T$ ($T$ < 357 K) or low-$P$ ($P$ < 0.5 GPa) insulating phase with the "S-Pbca" structure [22, 23]. Thus, it is anticipated that the application of $E$-fields to CRO is accompanied by the structural distortion to release the "flattening" (by the $c$-axis expansion).

To confirm the $E$-field-induced structural transition, we performed x-ray diffraction measurements for single-crystalline CRO in $E_{\parallel c}$ at 290 K. Figure 2 (a) shows comparison of the diffraction patterns, showing the (006) reflection at representative $E$-fields. The (006) peak at $E$ = 0 V/cm observed at 45.62 degree (11.915 Å) indicates that the system is in the stoichiometric S-Pbca insulating phase. In contrast, the application of $E$ turns the system from the S-Pbca phase with the short $c$-axis to the L-Pbca metallic phase with long $c$ (12.276 Å) via a mixed state where the metallic phase coexists with the insulating phase. Since the L-Pbca (006) peak becomes visible at ~ $E_{th}$, the switching is accompanied by a bulk first-order structural transition.

It should be noted that this structural transition is actually visible under a microscope. As evident in the supplemental video, contraction and expansion of a CRO crystal (showing its $ab$ plane) is induced by repeatedly applying "tiny electric pulses" of 25 mJ (100 ms, 8.5 V and 30 mA) at 290 K. This phenomenon may remind us of the piezoelectric effect in ferroelectric crystals but is actually due to the $I$-$M$ switching accompanied by a structural transition. We note that a large current often destroys the samples. However, the switching phenomenon driven by the tiny electric pulses can be stably repeated at least three thousand times. The *E*-field-induced Mott transition thus occurs in the bulk of the sample, not just on the surface or in a filamentary region.

As shown in Fig. 2 (b) and (c), the detailed process of the switching accompanied by a structural transition is obtained by a simultaneous measurement of volume fraction (b) and *I* (c) as functions of *E*. As seen in this comparison, with increasing *V* the switching precedes the structural transition. Moreover, the mixed phase persists in the field range from 36 to 48 V/cm. We can, thus, deduce that the bulk metallic region is initially formed along the *E* direction and then spreads through the whole crystal.

We note that an isovalent substitution of Sr for Ca [26] or pressurisation [13] turns a Mott insulator to a quasi-two-dimensional Fermi-liquid metal without any carrier doping. In these cases, the Mott transition is interpreted in terms of a switching in the orbital occupation driven by the lattice flattening distortion [27]. In contrast, application of an *E*-field itself cannot directly act on the structural distortions, and in fact, the switching occurs prior to the structural transition as indicated by the comparison between Fig. 2(b) and (c).

Let us next compare the lattice parameter *c* among the *E*-field, *P*, and heating induced structural transitions. Figures 2 (d - f) show the lattice parameter *c* as functions of *E*-field, *P*, and *T*, respectively. With increasing *E*-field, the lattice parameter $c \sim 11.92$ Å (S-Pbca) gradually increases, and reaches ~12.01 Å at $E_{th} \sim 40$ V/cm, where it abruptly changes to $c \sim 12.28$ Å of the L-Pbca phase in a narrow mixed-phase region. Clearly, there are quantitative similarities among the

$E$-field, $P$, and $T$ variations of the $c$ parameter in the vicinity of the Mott transitions.

However, we note here an important difference between the $E$-field and heating induced metallic states. Once the switching occurs, the lattice parameter $c \sim 12.28$ Å remains almost constant in the $E$-field range from 40 to 70 V/cm. In contrast, heating makes the $c$-parameter increase linearly at a rate of $\sim 5 \times 10^{-4}$ Å/K up to ~640 K. This adds further evidence that the $c$-parameter variation in the $E$-field cannot be identified as due to heating.

There remains an absorbing question: what is happening in the $E$-field-induced metallic phase at low temperatures? To the best of our knowledge, there has been no previous example of cooling a volatile switching-system while keeping the metallic state. With the parameters found, it would reach quite a large value, over 100 kV/cm at 4.2 K. In reality, it is extremely difficult to induce a Mott transition by the $E$-field at low temperatures.

Keeping this in mind, let us now present the results of $I$-biased experiments. Figure 3 (a) shows in-plane resistance measured with a constant current of 420 mA by a two-probe method as a function of $T$. Surprisingly, the $T$-variation shows a positive slope ( $d\rho_{ab}/dT > 0$ ) indicating metallic conduction in the $T$ range from 300 to 4.2 K. That is, once the switching has occurred, the $E$-field-induced metallic state becomes stable with flowing current even in $E$-fields much less than $E_{th}$. A heuristic analogy may be drawn with a well-known phenomenon that "flowing" suppresses the freezing point of water.

Figure 3 (b) shows the same data as (a) but for temperatures below 30 K. An abrupt change in the metallic slope at ~15 K is reminiscent of the resistivity change associated with a FM transition in the *P*-induced metallic CRO in Fig. 3 (c). Thus, we naturally anticipate that a FM ordered state appears also in this current-driven metal as a stationary but nonequilibrium state. Indeed, we have observed a change in the local magnetic field with magnetic probes. Although such measurements are technically not easy, they are in progress.

We have experimentally shown that a number of unusual phenomena emerge in the Mott insulator CRO by applications of electric fields and currents. First, the switching is induced by application of such low fields as $E_{th}$ ~40 V/cm, which is $10^2$~$10^3$ times lower than that reported in other Mott insulators. Second, the switching is accompanied by a bulk structural phase transition. Third, the induced metallic phase becomes stable to low temperatures if the flowing bias current is maintained. To add, we also show that under bias current the AFM transition, characteristic of the Mott insulating state of CRO, disappears and moreover FM ordering emerges below 15 K as indicated by the resistance drop as well as the spontaneous magnetization at low temperature.

**Discussion**

Now, let us start with a discussion of why CRO shows switching with a structural change. It might seem that switching at a low field is connected with "avalanche breakdown", a dielectric breakdown propagating from a small region of a sample associated with an impurity or surface effect. However, the switching in CRO cannot be understood as such an avalanche breakdown for the following reasons: first, the values of $E_{th}$ indicate good reproducibility, second, the switching characteristics are nearly independent of environmental atmosphere, and last, the switching is accompanied by a bulk structural change.

Among the intrinsic mechanisms, the observed $E_{th}$ ~40 V/cm is too small to be accounted for by a conventional Zener breakdown model. We should note that in a Mott insulator like CRO, for which conduction is frozen by Coulomb repulsion, unconventional mechanism of dielectric breakdown based on "the many-body Zener effect" is expected to be essential. In fact recent theoretical studies in terms of nonequilibrium processes in highly-correlated electron systems predict that a metallic state is induced at a threshold voltage much smaller than the Mott gap [28]. Such unconventional mechanism may explain relatively small threshold voltage of switching in some Mott insulators even without active orbital degeneracy [8-12].

In addition, orbital depolarization characteristic of a multiband system such as CRO may well be important in further reducing $E_{th}$. Specifically, it is known that in the insulating phase the lower orbital, Ru-$d_{xy}$, is fully occupied and the other bands, derived from the $d_{xz}$ and $d_{yz}$ orbitals, are half-filled, whereas in the metallic phase more electrons in the $d_{xz}$ and $d_{yz}$ lead to full orbital depolarisation. Such orbital depolarisation is strongly coupled to the structural transition. Due to the availability of this self-doping mechanism associated with spatial charge redistribution, the application of $E$-field may further enhance the electronic instability driven by the many-body Zener effect.

Here, we should also mention that the formula which describes the observed threshold energy to induce the metallic state, $E_{th}(T)/E_0 = \exp(-T/T_0)$, is often used for CDW systems and interpreted in terms of weakening of the pinning potential [11, 29]. This formula is importantly different from that for a thermal activation process, $E_{th}(T)/E_0 = \exp(T_0/T)$ and explains why $E_{th}$ is much smaller than the Mott gap at RT and why the *E*-field induced metal is maintained by current, in terms of delocalisation of carriers. A charge depinning process deduced from the observed behaviour suggests that the switching phenomena may be related to a ferro-type orbital ordering in the insulating phase [30], expected to be accompanied by charge localisation.

Lastly, it should be emphasized that the "flowing current" plays a key role in maintaining the induced metallic state, although the switching itself is induced not by the application of current but voltage. This phenomenon is truly unexpected and most probably requires a mechanism of stabilising a steady but nonequilibrium state (current maintained metal) over an equilibrium state (Mott insulator). Thus, application of *E*-fields and/or flowing currents has a great potential as a new tool to induce novel quantum phenomena in a variety of materials. Utilising such nonequilibrium states will certainly help expanding our knowledge of material science [31,32].

**Methods**

Single crystals of CRO were grown by a floating-zone method with $RuO_2$ self-flux. We used high quality samples [13-15] with essentially stoichiometric oxygen content judging from the *c*-axis lattice parameter of ~11.92 Å at 290K and the AFM transition at 113 K. Judging from the *c*-axis lattice parameter before and after switching, there is little change in the oxygen content during our switching experiments. Voltage-current (*V-I*) curves have been obtained with a load resistor (0~1 kΩ) connected in series and a dc voltage and current source/monitor (ADVANTEST, model TR6143 and ADCMT, model 6241A). The electrodes were made by gold evaporation on a freshly cleaved surface. The distance between the electrodes is typically 0.2 mm. We monitored a change in *T* at the sample during voltage sweeps. There exists little evidence for *T* rise in the *E* range below $E_{th}$. The resistance have often been measured by using a two-terminal method when there is negligible deference between a two- and a four-terminal one as shown in Fig. 1(e-g).

The x-ray diffraction study was performed for single-crystalline CRO in *E* applied along the *c* axis at 295 K. We irradiate the cleaved (001) surface with a ring-shaped electrode with x-ray. The contribution from $CuK\alpha_2$ radiation was analytically subtracted.

We have tried to examine a change in the local magnetic field directly using a superconducting quantum interference device (SQUID) magnetometer as well as a Hall probe. Detailed study is underway to determine the magnetic properties as a function of temperature.

**Acknowledgements**

We acknowledge T. Takemoto and Y. Kimura for their experimental helps, I. Terasaki and I. H. Inoue for fruitful discussions, and T. Yamagishi, S. Yonezawa at Kyoto University and R. Okazaki at Nagoya University for reproducing the result. We also acknowledge S. R. Julian for careful reading of the manuscript. A part of this work has been supported by a Grant-in-Aid for Scientific Research on Priority Areas (Grant No. 20029017 and 22540368) and by a Global COE grant "the Next Generation of Physics, spun from Universality and Emergence" from the MEXT of Japan.

**Author contributions**

F.N. planned the experiment and grew the samples. F.N., Y.M., S.T. and Y.Y. measured the electrical transport properties. M.S. and T.S. carried out the X-ray diffraction experiment. F.N. and Y.M. wrote the manuscript with contributions from the other authors.

**Author Information**

Correspondence and requests for materials should be addressed to F.N. (fumihiko@hiroshima-u.ac.jp).

**Additional Information**

**Competing financial interests:** The authors declare no competing financial interests.

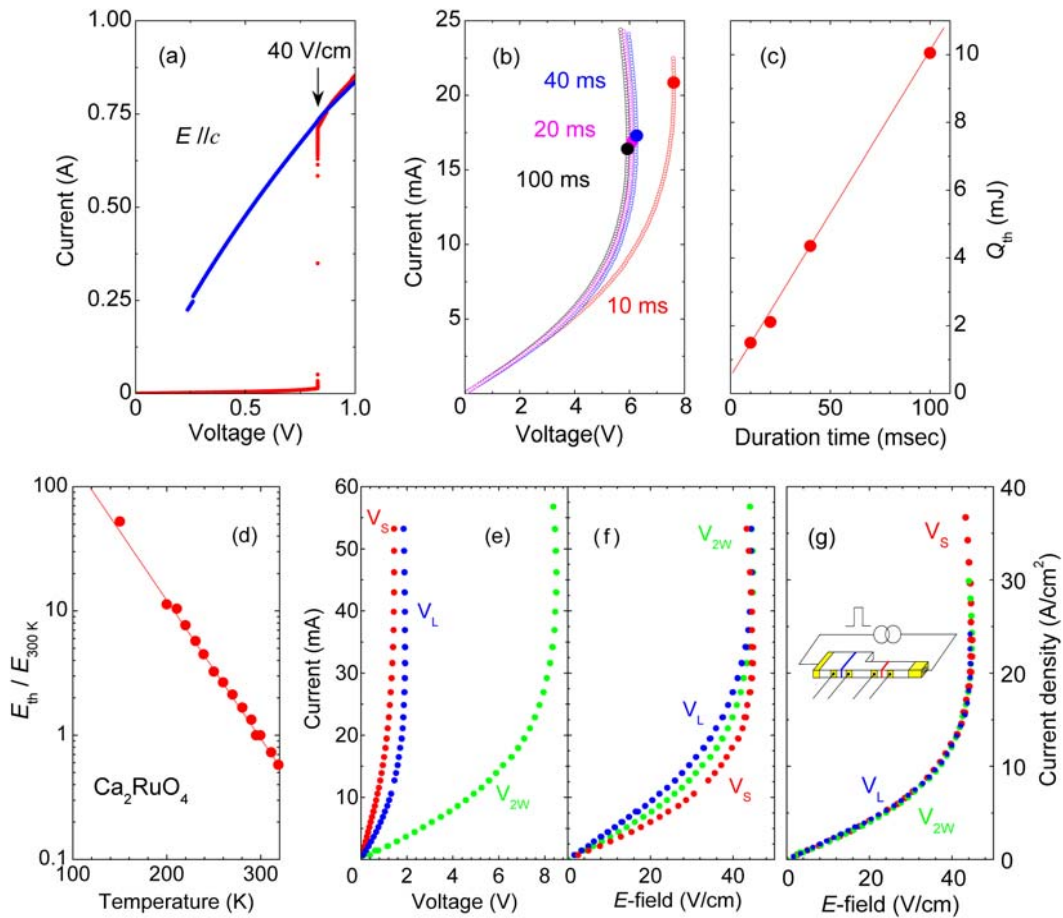

**Figure 1| Switching in voltage-current curves for $Ca_2RuO_4$.** (a) The voltage-current curves obtained by a two terminal method with continuous $E \parallel c$ at 295 K. There exists a large hysteresis during the voltage sweeps. (b) The switching curves for pulse application $E \perp c$ with a different duration time. The threshold $V_{th}$ and $I_{th}$ are defined from the maximum voltage in the *IV* curve. (c) Total heating $Q_{th}$ at the switching threshold estimated by $Q_{th} = \int V_{th}(t) \cdot I_{th}(t) \, dt$ in an adiabatic model plotted as a function of duration time. The almost linear increase of $Q_{th}$ with duration time gives clear evidence that the switching is not dominated by heating. The solid line is guide for the eye. (d) $E_{th}$ below 320 K plotted as a function of *T*. The solid line is a fit with $E_{th}(T)/E_0 = \exp(-T/T_0)$, using the characteristic values $E_0$ = 80 kV/cm and $T_0$ = 39.2 K. (e, f, g) The voltage-current curves measured by using a four-probe method for a step-shaped sample consisting of different cross-sectional areas. The schematic view of the step-shape sample is shown in the inset. The current are plotted against (e) voltage and (f) *E*-field. (g) The current density plotted as a function of *E*-field, showing that all the switching curves agree with each other.

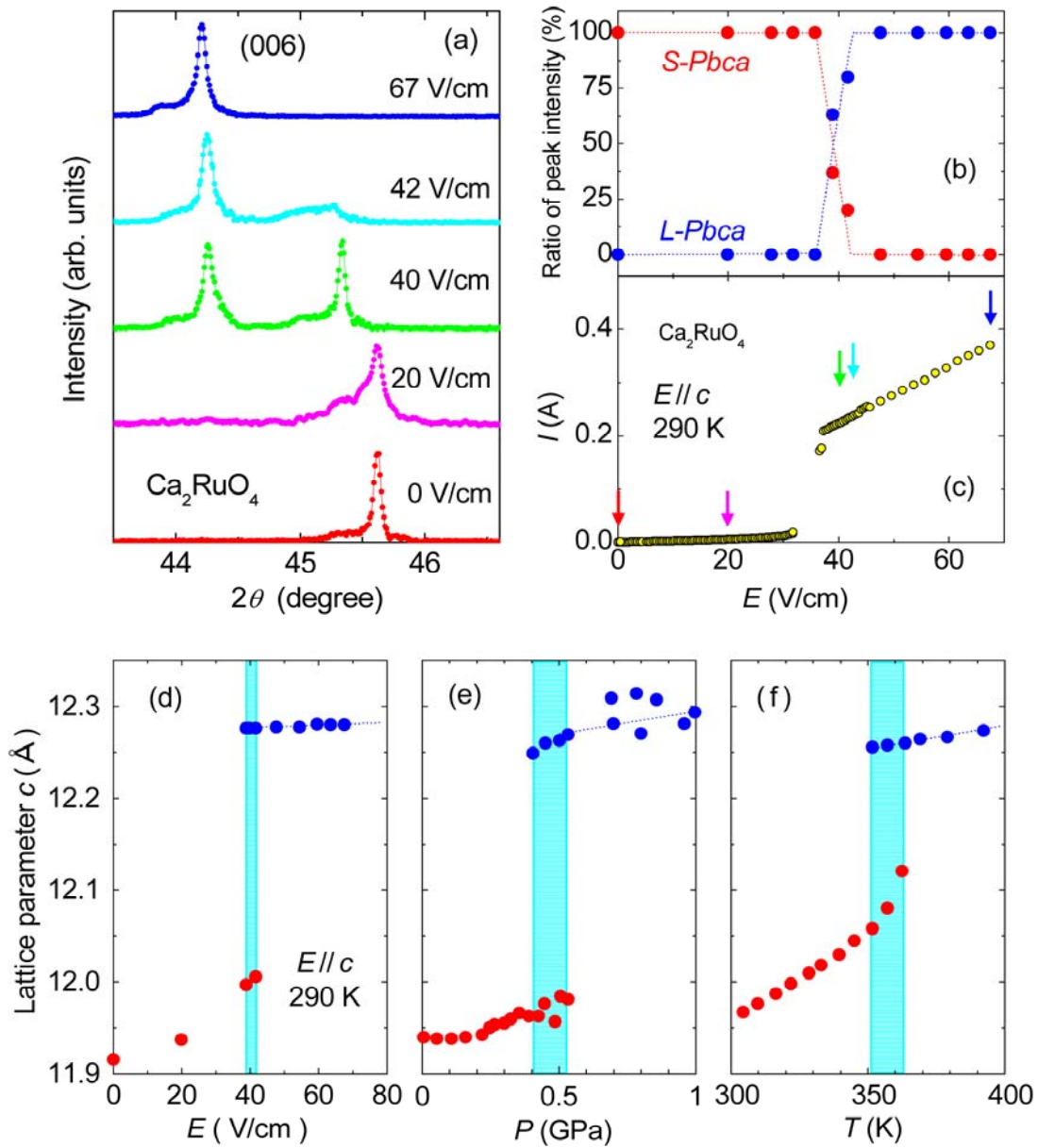

**Figure 2 | An x-ray diffraction study on single-crystalline $Ca_2RuO_4$ in electric fields applied along the *c* axis at 290 K.** (a) Comparison of the diffraction patterns, showing the (006) reflections at representative electric fields. (b) Volume fraction of S and L-Pbca phases plotted as a function of *E*. (c) Switching transition measured simultaneously with the x-ray diffraction. Comparing Fig. 2 (b) with 2 (c), we see that switching occurs prior to the structural transition. (d) Electric-field variation of the lattice parameter *c*. The phase transition from S to L-Pbca phase occurs at $E_{th}$ via the mixed phases (hatched region). (e) Pressure variation of lattice parameter *c*, from P. Stefen *et al*. (ref. 23). (f) Temperature variation of the lattice parameter *c*, from O. Friedt *et al*. (ref. 22).

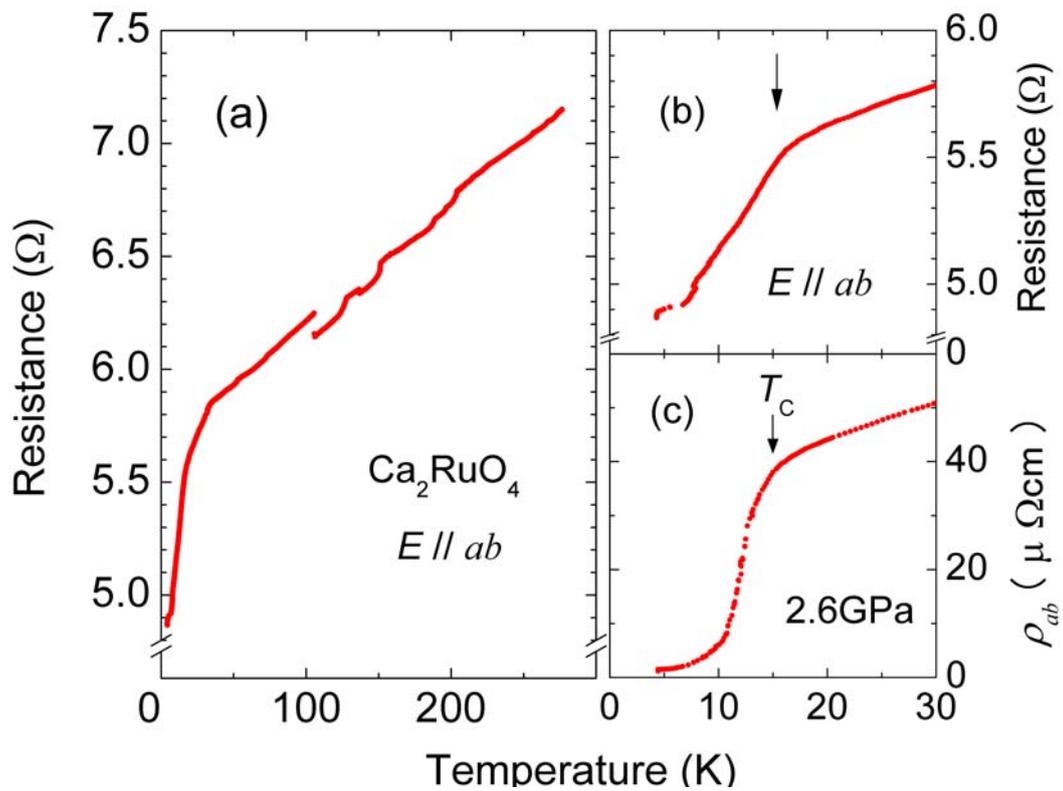

**Figure 3 | Temperature variations of in-plane resistance.** (a) With constant current $I$ = 420 mA. (b) Same data as (a) but for $T$< 30 K: The slope of the in-plane resistance in the electric field changes abruptly at ~15 K as indicated by an arrow. (c) In-plane resistivity under $P$ ~2.6 GPa, for which a similar variation to the $E$-field-induced case is seen at a temperature corresponding to ferromagnetic $T_C$.